\pdfoutput=1

\documentclass[11pt]{article}

\usepackage[preprint]{acl}

\usepackage{times}
\usepackage{latexsym}
\usepackage{xcolor}
\usepackage{booktabs}
\usepackage{longtable}

\usepackage[T1]{fontenc}

\usepackage[utf8]{inputenc}

\usepackage{microtype}

\usepackage{inconsolata}

\usepackage{graphicx}

\usepackage[linesnumbered,algoruled,boxed,lined]{algorithm2e}
\usepackage{multirow}
\usepackage{multicol}
\usepackage{booktabs}
\usepackage{epsfig}
\usepackage{algpseudocode}
\usepackage{subcaption}
\usepackage{amsmath} 
\usepackage{amsfonts}
\usepackage[para,online,flushleft]{threeparttable}

\newcommand{\Autoref}[1]{%
  \begingroup%
  \def\chapterautorefname{Chapter}%
  \def\sectionautorefname{Section}%
  \def\subsectionautorefname{Subsection}%
  \def\subsubsectionautorefname{Subsubsection}%
  \def\paragraphautorefname{Paragraph}%
  \def\tableautorefname{Table}%
  \def\equationautorefname{Equation}%
  \def\algorithmautorefname{Algorithm}%
  \autoref{#1}%
  \endgroup%
}

\title{MA-DPR: Manifold-aware Distance Metrics for Dense Passage Retrieval}

\author{Yifan Liu\thanks{Equal contribution}\textsuperscript{\textnormal{1}},
  Qianfeng Wen\footnotemark[1]\textsuperscript{\textnormal{1}}, Mark Zhao\footnotemark[1]\textsuperscript{\textnormal{1}}, 
  Jiazhou Liang\textsuperscript{\textnormal{1}},
  Scott Sanner\textsuperscript{\textnormal{1,2}}\\
  \textsuperscript{1}University of Toronto, Canada\\
  \textsuperscript{2}Vector Institute of Artificial Intelligence, Toronto, Canada\\
  \texttt{\{yifanliu.liu, qianfeng.wen, mark.zhao, joe.liang\}@mail.utoronto.ca} \\ \texttt{ssanner@mie.utoronto.ca}
}

\begin{document}
\maketitle
\begin{abstract}
Dense Passage Retrieval (DPR) typically relies on Euclidean or cosine distance to measure query-passage relevance in embedding space, which is effective when embeddings lie on a linear manifold.  However, our experiments across DPR benchmarks suggest that embeddings often lie on lower-dimensional, non-linear manifolds, especially in out-of-distribution (OOD) settings, where cosine and Euclidean distance  fail to capture semantic similarity. 
To address this limitation, we propose a \textit{manifold-aware} distance metric for DPR (\textbf{MA-DPR}) that models the intrinsic manifold structure of passages using a nearest neighbor graph and measures query-passage distance based on their shortest path in this graph. We show that MA-DPR outperforms Euclidean and cosine distances by up to 26\% on OOD passage retrieval with comparable in-distribution performance across various embedding models while incurring a minimal increase in query inference time. Empirical evidence suggests that manifold-aware distance allows DPR to leverage context from related neighboring passages, making it effective even in the absence of direct semantic overlap. MA-DPR can be applied to a wide range of dense embedding and retrieval tasks, offering potential benefits across a wide spectrum of domains.
\end{abstract}

\section{Introduction}
\label{sec:intro}


Dense Passage Retrieval (DPR) \citep{dpr} operates on the principle that semantically similar queries and passages remain close within a learned \emph{dense} embedding space. By ranking passages based on their distances to the query, DPR aims to measure semantic query-passage relationships rather than \emph{sparse} word-level matches.
DPR approaches primarily rely on Euclidean and cosine distances  \citep{mussmann2016learning, MIPS} due to their computational efficiency and straightforward interpretability. 

\begin{figure}
    \centering
    \includegraphics[width=1\linewidth]{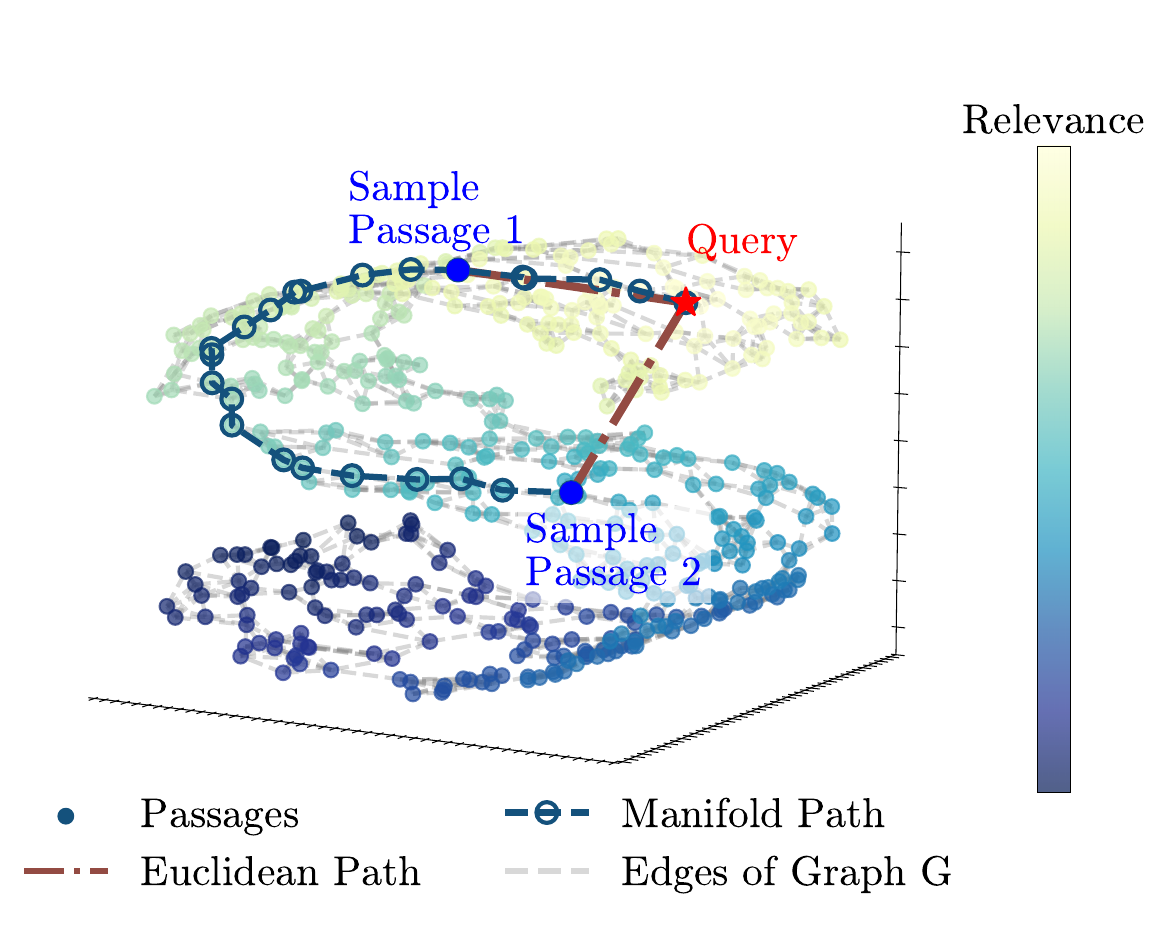}
    \caption{Example of subdimensional non-linear manifold in embedding space. Passage embeddings (dots) form a non-linear S-shaped manifold, where their relevance (indicated by color) to the query (red star) is determined by proximity along the manifold rather than Euclidean distance. Two sample passages (blue dots) have similar Euclidean distance to the query (red path) but differ in relevance. In contrast, the distance (blue path) along the weighted undirected graph $G$, where nodes represent passages and edges (gray dashed lines) connect each passage to its \( K \)-nearest neighbors, often better reflects the true relevance in such settings.}
    \label{fig:s-shape}
\end{figure}

However, the well-known manifold hypothesis states that high-dimensional data, such as text embeddings, reside on a subdimensional manifold \citep{tenenbaum2000isomap, roweis2000lle}. In such a case, relying solely on Euclidean and cosine distances may fail to capture the true relevance between queries and passages within this subdimensional manifold structure (cf. \autoref{fig:s-shape}).




\begin{figure*}[h!]
    \centering
    \includegraphics[width=0.95\linewidth]{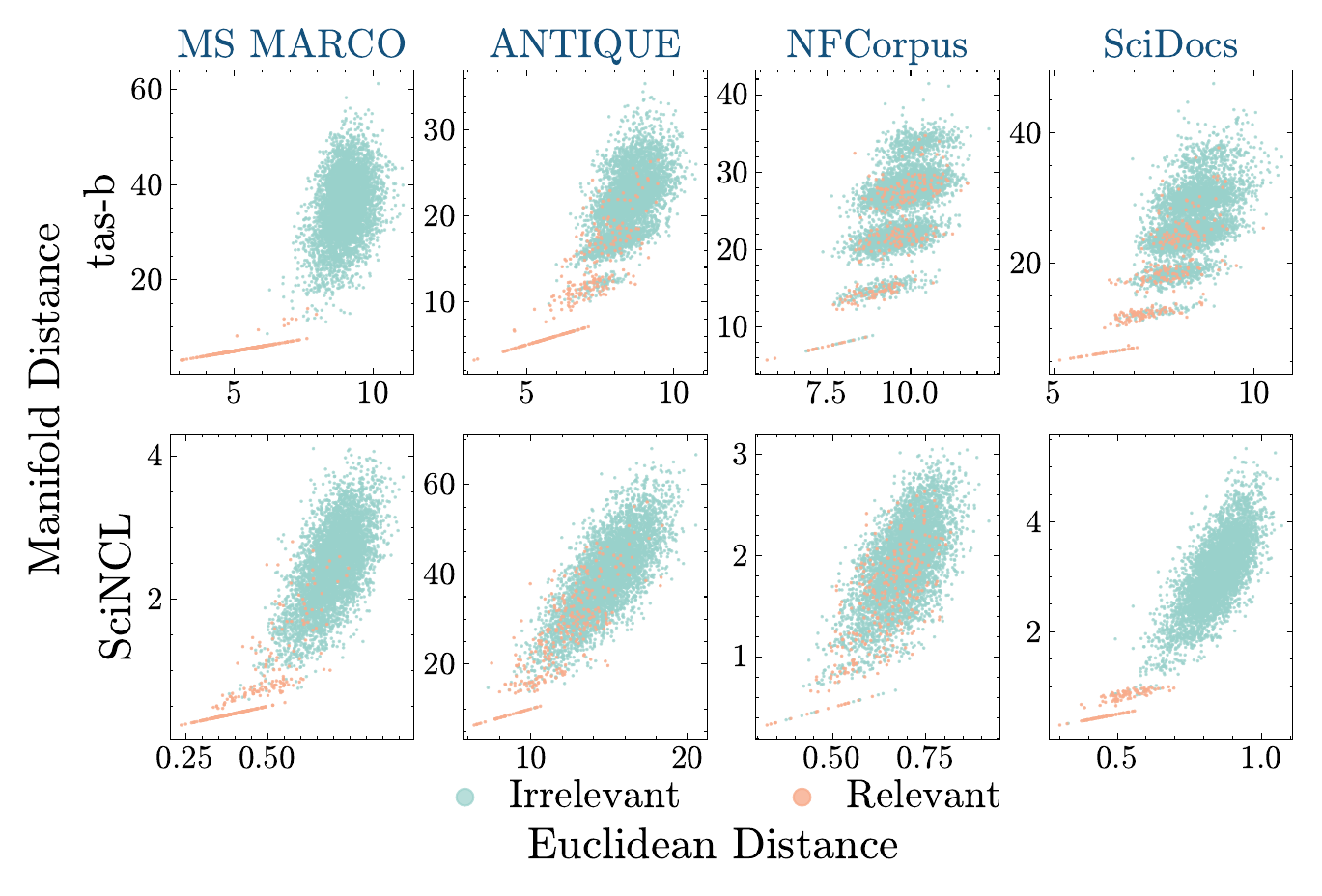}
    \caption{We evaluate the alignment between Euclidean distance (x-axis) and a manifold-aware distance (y-axis) for all query-passage pairs (relevant=orange, irrelevant=cyan) in the embedding space across four benchmark datasets (MS MARCO, ANTIQUE, NFCorpus, SciDocs) and two embedding models (\texttt{tas-b} and \texttt{SciNCL}), as detailed in \Autoref{sec:experiments}. In-distribution pairs (MS MARCO for \texttt{tas-b}, SciDocs for \texttt{SciNCL}) exhibit strong agreement and relevance distinction using both distance metrics (orange in lower left). The remaining OOD settings show more misalignment, where manifold distance sometimes offers improved relevance distinction over Euclidean distance.
    Two artifacts are noteworthy: (i) The orange ``line'' in the lower left is due to relevant documents that are 1-hop away from the query in the manifold graph, hence manifold distance equals Euclidean distance. (ii) The disconnected ``blobs'' present in many plots correspond to different numbers of hops from the query in the manifold graph.}

    \label{fig:distances_EuclideanvsManifold}
\end{figure*}

Dimensionality reduction methods \citep{jolliffe2002principal, scholkopf1997kernel} and metric learning methods \citep{goldberger2004neighbourhood, weinberger2009distance}
aim to learn a mapping from the original embedding space into a lower-dimensional representation that better reflects the subdimensional manifold structure. In such a space, Euclidean and cosine distances may better capture the true relevance between queries and passages. However, our empirical results show that these methods do not consistently improve DPR performance (cf. \Autoref{tab:experiment_results}), suggesting the need for alternative approaches to capture the manifold structure.

\textit{Manifold-aware} distance metrics construct a graph-based representation of the data \citep{belkin2003laplacian,tenenbaum2000isomap} that may better capture the (\textit{potentially non-linear}) structure of the subdimensional manifold than the dense embedding space, and then measure distances 
on this manifold. While earlier work has acknowledged the presence of
manifold structures in embedding spaces \citep{zhou2003ranking, yonghe2019refining, zhao2020manifold}, these important insights have been largely unexplored in the context of DPR. 

We hypothesize that 
manifold-aware distances better capture query-passage relevance in DPR when embeddings reside on a non-linear manifold and thus 
summarize our contributions as follows:  


\begin{enumerate}
\item We empirically verify the existence of subdimensional manifolds with non-linear structure in out-of-distribution (OOD) embedding spaces (cf. \Autoref{fig:distances_EuclideanvsManifold}), which motivates our investigation of manifold-aware distances.

\item We investigate a large design space of manifold-aware distance metrics for DPR (\textbf{MA-DPR}) that leverage a graph-based representation to exploit the non-linear manifold structure of the embedding space. 

\item We show that MA-DPR matches or outperforms DPR with Euclidean and cosine distances across benchmark datasets and embeddings with minimal impact on retrieval time. 
By leveraging neighboring passage context, we show MA-DPR can even retrieve relevant query-passage pairs lacking semantic overlap.
\end{enumerate}

\section{Related Work}
\label{sec:related_work}
\subsection{Dense Passage Retrieval}
\label{sec:dpr lit}
Dense Passage Retrieval (DPR) \citep{dpr} encodes queries and passages into a shared dense embedding space and ranks passages based on Euclidean and cosine distances \citep{mussmann2016learning, MIPS}. These distance metrics are widely used due to their computational efficiency, interpretability, and compatibility with scalable approximate nearest neighbor search methods \citep{MIPS}. The underlying assumption is that semantically relevant queries and passages are positioned close to each other in the embedding space.

However, this assumption is often violated, particularly in out-of-distribution (OOD) settings where embeddings are not optimized for the target domain. In such cases, Euclidean and cosine distances may fail to capture complex semantic relationships, leading to retrieval failures despite close proximity in dense embedding space \citep{steck2024cosine}. 

\subsection{Mapping \& Metric Learning Approaches}
\label{sec:mapping_and_metric_learning}
Given the limitations of Euclidean and cosine distances in the dense embedding space, a common solution is to \emph{learn a mapping} that transforms the original space into a new representation space. Classical dimensionality-reduction methods such as PCA \citep{jolliffe2002principal} and Kernel PCA \citep{scholkopf1997kernel} map embeddings into lower-dimensional representations that retain dominant variation. 

Metric learning methods such as NCA \citep{goldberger2004neighbourhood} and LMNN \citep{weinberger2009distance} learn supervised linear transformations that pull semantically related items closer and push dissimilar ones apart. These mapping methods offer parametric approximations to the underlying lower-dimensional manifold within the dense embedding space.

\subsection{Manifold-Aware Approaches}
 The manifold hypothesis posits that high-dimensional data, such as text embeddings, reside on subdimensional manifolds \citep{tenenbaum2000isomap, roweis2000lle}. This makes manifold-aware distance metrics a more natural solution for measuring relationships on both linear and non-linear manifolds, compared to standard Euclidean and cosine distances that make an implicit linear manifold assumption.

Previous studies, such as Isomap~\citep{tenenbaum2000isomap}, Locally Linear Embedding (LLE)~\citep{roweis2000lle}, and Laplacian Eigenmaps~\citep{belkin2003laplacian}, approximate manifold-aware distances by constructing neighborhood graphs or computing spectral embeddings that capture the underlying manifold structure.

Similar manifold-aware approaches have been applied in Information Retrieval (IR) to incorporate global structural information, particularly in image retrieval tasks where the embedding space more naturally adheres to manifold structures~\citep{zhou2003ranking, he2004manifold, yang2013saliency}. However, manifold-aware distance metrics have not been systematically explored in DPR, leaving a critical yet unaddressed gap in the current retrieval framework.


\section{MA-DPR: Manifold-aware Distance for Dense Passage Retrieval} \label{sec:method}
Let \( \mathcal{P} = \{p^{(1)}, p^{(2)}, \ldots, p^{(N)}\} \) denote a collection of \( N \) passages, each associated with a dense embedding in a \( D \)-dimensional space, denoted as \( \{\mathbf{e}_p^{(1)}, \dots, \mathbf{e}_p^{(N)}\} \), where \( \mathbf{e}_p^{(i)} \in \mathbb{R}^{D} \).

DPR ranks passages in \( \mathcal{P} \) by measuring their semantic similarity to a given query \( q \), computed as the distance between the query embedding \( \mathbf{e}_q \in \mathbb{R}^D \) and each passage embedding \( \mathbf{e}_p^{(i)} \) as:
\begin{equation}
\text{Rank}(q, \mathcal{P}) = \operatorname*{argsort}_{p^{(i)} \in \mathcal{P}} \, d(\mathbf{e}_q, \mathbf{e}_p^{(i)}),
\end{equation}
where \( d(\cdot, \cdot) \) is a defined distance metric. In practice, common choices for $d$ in DPR include Euclidean distance:
\begin{equation}\label{eq:euclidean}
d_{\text{Euclidean}}(\mathbf{e}_q, \mathbf{e}_p^{(i)}) = \left\|\mathbf{e}_q - \mathbf{e}_p^{(i)}\right\|_2,
\end{equation}
and cosine distance:
\begin{equation} \label{eq:cosine}
d_{\text{Cosine}}(\mathbf{e}_q, \mathbf{e}_p^{(i)}) = 1 - \frac{\mathbf{e}_q \cdot \mathbf{e}_p^{(i)}}{\left\|\mathbf{e}_q\right\|_2 \left\|\mathbf{e}_p^{(i)}\right\|_2}.
\end{equation}

Both $d_{\text{Euclidean}}$ and $d_{\text{Cosine}}$ assume that semantic similarity corresponds to proximity in the dense embedding space, but this assumption can fail to capture true query–passage relationships under subdimensional manifold structures, as discussed in \Autoref{sec:dpr lit}. 

Thus, we propose MA-DPR, an extension of DPR with a manifold-aware distance metric $d_{\text{Manifold}}$ to overcome the existing limitation. MA-DPR operates in two stages: (1) a one-time \textit{offline} manifold graph construction (cf. \Autoref{sec:graph_construction}) to capture the subdimensional manifold structure within embedding space, followed by (2) an \textit{online} passage ranking stage for a given query based on the constructed graph (cf. \Autoref{sec:passage_ranking}).





\subsection{Manifold Graph Construction} \label{sec:graph_construction}
We propose a weighted undirected graph \( G \) to approximate the manifold structure of the embedding space (cf. \Autoref{fig:s-shape}) for $\mathcal{P}$. Each passage \( p^{(i)} \in \mathcal{P} \) is represented as a vertex, and edges connect \( p^{(i)} \) to its \( K \)-Nearest Neighbors (KNN) based on their proximity in the embedding space. 


Specifically, let \( G = (V, E, c) \), where:
\begin{description}
    \item[\( V \):] A set of vertices $\{v_1,\dots,v_N\}$  representing $p^{(i)}, i\!\in\!\{1,\dots,N\}$.

    \item[\( E \):] A set of edges, where an edge \( \{v_i, v_j\} \) exists between two vertices $v_i$ and $v_j$ if either $\mathbf{e}^{(i)}_p$ or $\mathbf{e}^{(j)}_p$  is among the KNN of the other based on a defined distance metric \( d^{\text{KNN}} \). 
    
    \item[\( c \):]  A cost function \( c(v_i, v_j) : E \to \mathbb{R} \) assigns a cost to $\{ v_i, v_j \}$.
\end{description}

Before introducing the passage ranking stage, we first present the proposed design choices for constructing the manifold graph: (1) the local distance metric~\(d^{\text{KNN}}\) used to identify the \(K\)-nearest neighbors, (2) the edge cost function~\(c\), and (3) the number of nearest neighbors~\(K\).

\subsubsection{Choices of Distance metrics \( d^{\text{KNN}} \)}
\label{distance_function}
\paragraph{Euclidean and cosine distances}
When local neighbor distances are informative, as illustrated in \Autoref{fig:s-shape}, the distance metric used in DPR can serve as an intuitive choice of \( d^{\mathrm{KNN}} \) for KNN graph construction. Common choices include the Euclidean distance \( d^{\mathrm{KNN}}_{\text{Euclidean}} \) (cf. \Autoref{eq:euclidean}) and the cosine distance \( d^{\mathrm{KNN}}_{\text{Cosine}} \) (cf. \Autoref{eq:cosine}).

However, as illustrated in \Autoref{sec:dpr lit}, such distances in embedding space are not always reliable: $d^{\text{KNN}}_{\text{Euclidean}}$ and $d^{\text{KNN}}_{\text{Cosine}}$ may fail to capture the 
underlying semantic relationship between local neighbors. Moreover, it is sensitive to noise and local perturbations.


\paragraph{Spectral distance}
A widely adopted approach for $d^{\text{KNN}}$ is to compute the distance between passages in a \textit{spectral embedding space}, leveraging the \textit{eigenstructure of the graph Laplacian}~\citep{shi2000normalized, belkin2003laplacian}.

Crucially, the spectral distance captures the intrinsic structure of the embedding manifold by leveraging global graph connectivity, rather than relying solely on local neighborhoods \citep{ploux2003semantic, dhillon2004unified, tsai2016multinomial}. This makes it well aligned with the manifold-aware objective of our work and inherently less sensitive to local noise or perturbations.

The \textit{spectral distance} between two passages \( p^{(i)} \) and \( p^{(j)} \) is defined as:
\begin{equation}
d^{\mathrm{KNN}}_{\mathrm{Spectral}}\left(\mathbf{e}_p^{(i)}, \mathbf{e}_p^{(j)}\right) = \left\|\mathbf{u}^{(i)} - \mathbf{u}^{(j)}\right\|_2,
\end{equation}
where \( \mathbf{u}^{(i)} \in \mathbb{R}^M \) denotes the \( m \)-dimensional spectral embedding of passage \( p^{(i)} \) obtained from the top \( m \) non-trivial eigenvectors of the normalized graph Laplacian. This embedding arises by minimizing 
\(\mathrm{Tr}(U^\top L_{\mathrm{sym}}U)=\tfrac{1}{2}\sum_{i,j} w_{ij}\|\tfrac{\mathbf{u}^{(i)}}{\sqrt{d_i}}-\tfrac{\mathbf{u}^{(j)}}{\sqrt{d_j}}\|_2^2\) 
subject to \(U^\top U=I\), whose solution is given by these eigenvectors.


\subsubsection{Choice of Cost Function $c$} \label{cost_function}
Once the KNN Graph $G$ is constructed as depicted in the graph mesh in \Autoref{fig:s-shape}, we need to define the cost of edges $E$ to compute distances during manifold-aware dense retrieval.


\paragraph{Distance Cost} Distance Cost (DC) directly utilizes $d^{\text{KNN}}$ as $c$: 
\begin{equation}
    c^{\text{DC}}(v_i, v_j) = d^{\text{KNN}}\left(\mathbf{e}_p^{(i)}, \mathbf{e}_p^{(j)} \right)
\end{equation}  
$c^{\text{DC}}$ preserves the original distance metric used in $d^{\text{KNN}}$ and serves as an intuitive choice for defining the cost function.


\paragraph{Uniform Cost} 
Uniform Cost (UC) assigns the same constant cost to every edge, independent of the embedding distance:  
\begin{equation}
    c^{\text{UC}}(v_i, v_j) = 1.
\end{equation}
UC emphasizes discrete connectivity by counting the number of edges (hops) along the shortest path, which mitigates the influence of local noise and potentially unreliable embedding distances in DC. 



\begin{algorithm}[!h]
\caption{Passage Ranking via manifold-aware distance}
\label{alg:inference-time}
\KwIn{Query embedding $\mathbf{e}_q$, passage embeddings $\{\mathbf{e}_p^{(1)}, \ldots, \mathbf{e}_p^{(N)}\}$, pre-constructed graph $G=(V,E,c)$, number of neighbors $K$, distance metric $d$.}
\KwOut{Ranked passages $R_K(q)$ with respect to query $q$.}

\textbf{Step 1: Query Integration}\\
Introduce new vertex $v_{N+1}$ for $q$; \quad
$V \gets V \cup \{v_{N+1}\}$\;
$\mathcal{N}(q) \gets \operatorname{KNN}\!\big(\mathbf{e}_q,\{\mathbf{e}_p^{(i)}\}_{i=1}^N,d\big)$\;
\For{$v \in \mathcal{N}(q)$}{
  $E \gets E \cup \{(v_{N+1}, v)\}$; \quad $c(v_{N+1}, v) \gets d(\mathbf{e}_q,\mathbf{e}_v)$\;
}

\textbf{Step 2: Passage Ranking}\\
Compute \(d_{\text{Manifold}}(q, v_i)\) from \(q\) to all passages \(v_i \in V\)\;
Return indices of the top-\(K\) passages\ based on the \(d_{\text{Manifold}}(q, v_i)\) in ascending order.

\textbf{Step 3: Cleanup}\\
$E \gets E \setminus \{(v_{N+1},u),(u,v_{N+1}) : u \in V\}$; \quad
$V \gets V \setminus \{v_{N+1}\}$\;

\end{algorithm}

\subsubsection{Choice of $K$}

Selecting the value of \(K\) for KNN involves a trade-off: \emph{smaller values of \(K\) emphasize local structure by connecting only to immediate neighbors, but may lead to disconnected graphs and sensitivity to noise and individual document perturbations}. 

\emph{In contrast, larger values of \(K\) reduce the influence of local noise and perturbations by emphasizing global connectivity and manifold smoothness, but could oversimplify the manifold structure by treating large neighborhoods as locally linear.}

\Autoref{distance_function} and  \Autoref{cost_function} jointly define a two-fold design space for constructing the graph \( G \). \Autoref{sec:experiments} presents empirical evaluations of each design choice (RQ3) and the impact of varying \( K \) on retrieval performance (RQ4) across a range of DPR benchmarks.


\subsection{Passage Ranking}
\label{sec:passage_ranking}
Given a constructed graph \( G = (V, E, c) \), the manifold-aware distance \( d_{\text{Manifold}} \) between two passages is defined as the minimum total edge cost along any path connecting the corresponding vertices \( v_i \) and \( v_j \) in \( G \) (a.k.a. the shortest path).

\( d_{\text{Manifold}} \) is formally defined as follows:
\begin{equation}
d_{\text{Manifold}}(v_i, v_j) = \min_{\pi \in \Pi(v_i, v_j)} \sum_{(u, v) \in \pi} c(u, v),
\end{equation}
where \( \Pi(v_i, v_j) \) denotes the set of all paths from \( v_i \) to \( v_j \), and \( \pi \in \Pi(v_i, v_j) \) is a specific path represented as a sequence of edges \( (u, v) \in E \).

To compute \( d_{\text{Manifold}} \) for a query \( q \) with embedding \( \mathbf{e}_q \), \( q \) is temporarily added to \( G \) as a new vertex \( v_{N+1} \). Edges are then formed by connecting \( v_{N+1} \) to its \( K \)-nearest passage vertices using defined $d^{\mathrm{KNN}}$ based on \( \mathbf{e}_q \).  

For MA-DPR, each passage \( p^{(i)} \) is ranked according to its manifold-aware distance from the query, \( d_{\text{Manifold}}(v_{N+1}, v_i) \), 
A smaller value indicates a shorter path along the manifold in the embedding space, signifying higher relevance, while a larger \( d_{\text{Manifold}}(v_{N+1}, v_i) \) implies lower relevance. The full procedure of passage ranking is described in \Autoref{alg:inference-time}.

\subsection{Complexity Analysis}
\label{sec:runtime}

MA-DPR introduces a one-time offline cost for constructing the manifold graph over the passage embeddings. This graph construction is independent of the query and does not affect the runtime efficiency of passage ranking at inference time.

At query time, the passage ranking process consists of two main steps: (1) computing distances between the query and all \( N \) passages to identify its \( K \)-nearest neighbors, with complexity \( \mathcal{O}(N D) \); and (2) computing manifold-aware distances via shortest-path traversal on the KNN graph using Dijkstra’s algorithm, which has complexity \( \mathcal{O}(K N + N \log N) \), given \( |E| = \mathcal{O}(K N) \). Thus, the per-query complexity of MA-DPR is:
\[
\mathcal{O}\!\left(N(D + K + \log N)\right).
\]

In a simulation with 100$k$ passages, we compared DPR and MA-DPR across embedding dimensions 
\(D \in \{32,64,128,256,512,1024\}\) and neighborhood sizes \(K \in \{2,\dots,15\}\). 
As shown in \Autoref{tab:runtime_comparison} and \Autoref{sec:runtime} (for full results), MA-DPR exhibits comparable per-query runtimes to DPR, 
with only a small overhead of about 2--4\,ms across dimensions. 
This confirms that incorporating manifold-aware distances preserves the efficiency of DPR at inference time 
while introducing negligible additional latency.

\begin{table}[h]
\centering
\resizebox{\columnwidth}{!}{
\begin{tabular}{lcc}
\toprule
\textbf{Method} & \textbf{Complexity} & \textbf{Latency (100K)} \\
\midrule
DPR & \( \mathcal{O}(ND) \) & 5.05 [5.04, 5.06] ms \\
MA-DPR & \( \mathcal{O}(N(D + K + \log N)) \) & 7.92 [7.73, 8.10] ms \\
\bottomrule
\end{tabular}
}
\caption{Per-query computational complexity and empirical latency of DPR (with \( d_{\text{Euclidean}} \)) and MA-DPR (with \( d^{\text{KNN}}_{\text{Euclidean}} + c^{\text{DC}} \)) for ranking over 100K passages. Shown here is the case of \(K=8\) and \(D=32\). Full results are reported in \Autoref{tab:runtime_madpr_by_k_with_dpr}. Results are reported with 95\% confidence intervals in $[\cdot]$\protect\footnotemark.}
\label{tab:runtime_comparison}
\end{table}

\footnotetext{System specifications: CPU—Intel(R) Core(TM) i7-14700HX 
; GPU—NVIDIA GeForce RTX 4070 Laptop GPU 
Average CPU utilization during measurement: \(\sim\)5\%.}






\section{Experiments}\label{sec:experiments}

All codes and results are available online\footnote{\href{https://github.com/QianfengWen/Manifold_Distance_Retrieval.git}{github.com/QianfengWen/Manifold\_Distance\_Retrieval.git}}.

\subsection{Experimental Setup} 


Our experiments aim to evaluate the effectiveness of MA-DPR \( d_{\text{Manifold}} \) against the following baselines as introduced in \Autoref{sec:related_work} and further described in mathematical detail in \Autoref{sec:baseline}:
\begin{itemize}
    \item DPR with \(d_{\text{Euclidean}}\)
    
    \item DPR with \(d_{\text{Euclidean}}\) + linear PCA
    \item DPR with \(d_{\text{Euclidean}}\) + Kernel PCA (Quadratic)
    \item DPR with \(d_{\text{Euclidean}}\) + Kernel PCA (RBF)
    \item DPR with \(d_{\text{Euclidean}}\) + NCA
\end{itemize}

Experiments are conducted on four standard DPR benchmarks:  
\begin{itemize}
    \item MS MARCO~\citep{nguyen2016ms}
    \item NFCorpus~\citep{boteva2016full}
    \item SciDocs~\citep{cohan2020specter}
    \item ANTIQUE~\citep{hashemi2020antique}
\end{itemize}

We report results using two embedding models:  
\begin{itemize}
    \item \texttt{msmarco-distilbert-base-tas-b} (\texttt{tas-b})~\citep{Hofstaetter2021_tasb_dense_retrieval}, trained on MS MARCO
    \item \texttt{SciNCL}~\citep{ostendorff2022neighborhoodcontrastivelearningscientific}, trained on SciDocs
\end{itemize}
The choice of embedding models naturally defines MS MARCO as the in-distribution dataset for \texttt{tas-b} and SciDocs as the in-distribution dataset for \texttt{SciNCL}, while the remaining datasets are treated as OOD. All embeddings are \(\ell_2\)-normalized.\footnote{\Autoref{sec:norm} provides additional empirical analysis on the performance of MA-DPR without normalization.}

For empirical evaluation, we assess the Recall, Mean Average Precision (MAP), and Normalized Discounted Cumulative Gain (nDCG) for the top 20 ranked assignments of each retrieval result.\footnote{The number of neighbors \( K \) in the KNN graph is fixed to 8 for all experiments except RQ4 (where it is varied).  The spectral embedding dimension \( M \) is fixed to 700. Evaluation of hyperparameters $K$ and $M$ is provided in \Autoref{sec:k}.}

Specifically, we address the following key research questions:  
\begin{itemize}
 \item[\textbf{RQ1:}] \textbf{Manifold Hypothesis Validation} Does empirical evidence support the presence of subdiensional manifold in dense embedding space? 
\item[\textbf{RQ2:}] \textbf{MA-DPR vs Baseline} Does MA-DPR lead to improved retrieval performance compared to other DPR baselines?
    
     \item[\textbf{RQ3}] \textbf{Design Choice Comparison} Which design choices for the manifold graph (cf. \Autoref{sec:graph_construction}) yield the best performance?  
        \item[\textbf{RQ4}] \textbf{Effect of K} What is the effect of varying the number of \( K \)-nearest neighbors in the manifold graph on the performance of MA-DPR?
            
    \item[\textbf{RQ5}] \textbf{Reasons for Improvement} For which queries can \( d_{\text{Manifold}}\) outperform \( d_{\text{Euclidean}} \)? What contributes to this improved performance?
\end{itemize}


  \subsection{Experimental Results}

\begin{table*}
\centering
\caption{Performance of MA-DPR across design choices on four datasets. The best results are highlighted in \textbf{bold}. An asterisk (*) denotes a statistically significant improvement of MA-DPR over DPR baselines (paired \( t \)-test, \( p < 0.05 \)). A slash (/) indicates results not reported. We normalize \texttt{tas-b} and \texttt{SciNCL} embeddings so that \( d_{\text{Euclidean}} \) and \( d_{\text{Cosine}} \) produce identical rankings.}
\resizebox{\textwidth}{!}{
\setlength{\tabcolsep}{3pt} 
\begin{tabular}{clccc|ccc|ccc|ccc}
\toprule
 & & \multicolumn{3}{c}{\textbf{NFCorpus}} 
 & \multicolumn{3}{c}{\textbf{SciDocs}}
 & \multicolumn{3}{c}{\textbf{ANTIQUE}} 
 & \multicolumn{3}{c}{\textbf{MS MARCO}} \\
\cmidrule(lr){3-5} \cmidrule(lr){6-8} \cmidrule(lr){9-11} \cmidrule(lr){12-14}
 &  & R@20 & mAP@20 & nDCG@20 & R@20 & mAP@20 & nDCG@20 & R@20 & mAP@20 & nDCG@20 & R@20 & mAP@20 & nDCG@20 \\
\midrule
\multicolumn{14}{c}{\centering msmarco-distilbert-base-tas-b} \\
\midrule
\multirow{3}{*}{DPR} 
 & 
 $d_{\text{Euclidean}}$ & 0.135 & 0.081 & 0.217 & 0.172 & 0.074 & 0.141 & 0.432 & 0.299 & 0.430 & 0.950 & 0.534 & 0.638 \\ 
  & 
 $d_{\text{Euclidean}} + \text{PCA}$ & 0.123 & 0.078 & 0.201 & 0.176 & 0.077 & 0.146 & 0.240 & 0.147 & 0.248 & \textbf{0.951} & \textbf{0.552} & \textbf{0.651} \\
  & 
 $d_{\text{Euclidean}} + \text{NCA}$ & 0.138 & 0.081 & 0.220 & / & / & / & 0.263 & 0.171 & 0.282 & \textbf{0.951} & 0.535 & 0.639 \\
 
 \addlinespace[8pt]

\multirow{4}{*}{MA-DPR} 
 & $d^{\text{KNN}}_{\text{Euclidean}}$+$c^{\text{UC}}$ & 0.143 & 0.083 & 0.222 & 0.182 & 0.076 & 0.146 & \textbf{0.467}* & 0.311 & 0.447 & 0.946 & 0.534 & 0.637 \\\addlinespace[2pt]
 & $d^{\text{KNN}}_{\text{Euclidean}}$+$c^{\text{DC}}$ & 0.137 & 0.083 & 0.220 & 0.167 & 0.074 & 0.139 & 0.417 & 0.299 & 0.424 & 0.945 & 0.534 & 0.637 \\\addlinespace[2pt]
 & $d^{\text{KNN}}_{\text{Spectral}}$+$c^{\text{UC}}$ & \textbf{0.147}* & \textbf{0.085} & 0.223 & \textbf{0.187}* & \textbf{0.078} & \textbf{0.148} & 0.464 & \textbf{0.317}* & \textbf{0.449}* & 0.944 & 0.534 & 0.636 \\\addlinespace[2pt] 
 & $d^{\text{KNN}}_{\text{Spectral}}$+$c^{\text{DC}}$ & \textbf{0.147}* & \textbf{0.085} & \textbf{0.228} & 0.182 & 0.077 & 0.146 & 0.436 & 0.307 & 0.430 & 0.932 & 0.498 & 0.604 \\
\midrule
\multicolumn{14}{c}{\centering SciNCL}  \\ 
\midrule
\multirow{3}{*}{DPR} 
 & $d_{\text{Euclidean}}$ & 0.119 & 0.073 & 0.191 & 0.275 & 0.116 & 0.215 & 0.239 & 0.140 & 0.228 & 0.622 & 0.251 & 0.339 \\ 
  & 
  $d_{\text{Euclidean}} + \text{PCA}$ & 0.115 & 0.070 & 0.183 & 0.260 & 0.112 & 0.206 & 0.111 & 0.060 & 0.116 & 0.639 & \textbf{0.254} & 0.346 \\
  & 
 $d_{\text{Euclidean}} + \text{NCA}$ & 0.100 & 0.078 & 0.157 & / & / & / & 0.131 & 0.074 & 0.124 & 0.626 & 0.250 & 0.341 \\
 
\addlinespace[8pt]

\multirow{4}{*}{MA-DPR} 
 & $d^{\text{KNN}}_{\text{Euclidean}}$+$c^{\text{UC}}$ & \textbf{0.133}* & \textbf{0.081}* & 0.203 & 0.266 & 0.116 & 0.211 & \textbf{0.250}* & \textbf{0.145} & \textbf{0.235} & \textbf{0.652}* & \textbf{0.254} & \textbf{0.348} \\\addlinespace[2pt]
 & $d^{\text{KNN}}_{\text{Euclidean}}$+$c^{\text{DC}}$ & 0.130 & \textbf{0.081}* & \textbf{0.205}* & \textbf{0.279} & \textbf{0.119} & \textbf{0.217} & 0.227 & 0.140 & 0.224 & 0.636 & 0.253 & 0.344 \\\addlinespace[2pt]
 & $d^{\text{KNN}}_{\text{Spectral}}$+$c^{\text{UC}}$ & 0.126 & 0.079 & 0.200 & 0.260 & 0.115 & 0.208 & 0.245 & 0.144 & 0.233 & 0.639 & 0.253 & 0.345 \\\addlinespace[2pt]
 & $d^{\text{KNN}}_{\text{Spectral}}$+$c^{\text{DC}}$ & 0.132 & \textbf{0.081}* & 0.204 & 0.260 & 0.112 & 0.204 & 0.233 & 0.140 & 0.225 & 0.623 & 0.246 & 0.335 \\
\bottomrule
\end{tabular}\label{tab:experiment_results}
}
\end{table*}

\paragraph{\textbf{RQ1 \textbf{Manifold Hypothesis Validation}:}}
To empirically validate the manifold hypothesis in dense embedding spaces, we examine the relationship between $d_{\text{Manifold}}$ and $d_{\text{Euclidean}}$ across relevant and irrelevant query-passage pairs.
Specifically, in \Autoref{fig:distances_EuclideanvsManifold}, for each ground truth relevant query \( q \) and passage \( p \) pair (orange dots) and irrelevant pair (blue dots), we compute \(d_{\text{Euclidean}}(q,p)\)  and  \(d_{\text{Manifold}}(q,p)\)  based on \(d^{\text{KNN}}_{\text{Euclidean}}\) + \(c^{\text{DC}}\) for manifold graph construction.

 In a perfectly linear embedding space, the manifold-aware distance induced by \( d^{\text{KNN}}_{\text{Euclidean}} + c^{\text{DC}} \) should closely align with standard Euclidean distance. However, in the presence of non-linear structure, the two distances are expected to diverge. This contrast enables us to diagnose and characterize non-linear relationships in the embedding space.
 
 Based on this intuition, we first observe a strong correlation between \( d_{\text{Euclidean}} \) and \( d_{\text{Manifold}} \) in scatterplots of relevant query-passage pairs on the in-distribution settings (i.e., MS MARCO w.r.t \texttt{tas-b} and SciDocs w.r.t. \texttt{SciNCL}). This correlation indicates that query-passage embeddings approximately lie on a locally linear manifold.
 
 However, in the remaining OOD settings---where embeddings were not directly optimized during training---\( d_{\text{Euclidean}} \) and \( d_{\text{Manifold}} \) exhibit significant misalignment, indicating that the embedding space follows a subdimensional non-linear manifold for both embedding models.

Further analysis reveals that in most OOD datasets, both relevant and irrelevant query-passage pairs fall within a similar range of \( d_{\text{Euclidean}} \), causing them to be ranked similarly in DPR. This supports our conjecture (cf. \Autoref{fig:s-shape}) that \( d_{\text{Euclidean}} \) 
fails to capture the true semantic relationship between query and passage. In contrast, \( d_{\text{Manifold}} \) more effectively separates relevant and irrelevant query-passage pairs, with relevant pairs consistently exhibiting smaller \( d_{\text{Manifold}} \) values. 
These findings suggest that \( d_{\text{Manifold}} \) could better capture query-passage relationships by modeling the intrinsic non-linear manifold structure of the embedding space, motivating our further investigation in \textbf{RQ2}.

\paragraph{\textbf{RQ2 \textbf{MA-DPR vs Baseline}:}} 
Motivated by \textbf{RQ1}, we empirically compare the performance of  MA-DPR and other DPR baselines in \Autoref{tab:experiment_results}.\footnote{For mapping-based baselines, we report only the best-performing mapping dimension and PCA kernel in \Autoref{tab:experiment_results}. We provide complete results in \Autoref{sec:baseline}. We do not report metric learning results for SciDocs, as the dataset lacks a separate training set required for supervised metric learning.}

We first observe that the mapping-based approach does not yield meaningful improvements over DPR with $d_{\text{Euclidean}}$ across datasets and embedding models, and even leads to performance decreases on NFCorpus and ANTIQUE. This suggests that such methods fail to adequately capture the non-linear manifold structure of the embedding space as empirically shown in RQ1, thereby motivating the exploration of alternative approaches.

Across nearly all four design choices, MA-DPR significantly outperforms baseline methods on OOD datasets without parameter tuning. This aligns with RQ1, where OOD datasets exhibit a subdimensional non-linear manifold structure. In such cases, \( d_{\text{Manifold}} \) more effectively captures the underlying data manifold structure and appears to better capture relevance.

In contrast, on in-distribution datasets, where the embedding space approximates a locally linear manifold, as empirically shown in RQ1, both MA-DPR and DPR baselines yield similar retrieval performance. In such cases, \( d_{\text{Manifold}} \) effectively reduces to \( d_{\text{Euclidean}} \), as the shortest path along the manifold aligns with the Euclidean distance (cf. \Autoref{fig:distances_EuclideanvsManifold}).

These results suggest that the manifold hypothesis validation shown in RQ1 offers a natural test for determining whether MA-DPR can be deployed on a new dataset. If the dataset exhibits a non-linear manifold structure, MA-DPR is expected to be effective and could be deployed. Otherwise, MA-DPR and DPR baselines are likely to yield similar performance.

In addition,  MA-DPR outperforms DPR with Euclidean and cosine distances across both embeddings, indicating that its effectiveness is not tied to a specific embedding space. This improvement underscores its robustness to variations in the embedding model, further validating its generalizability.

\paragraph{RQ3 \textbf{Design Choice Comparison}:}
\Autoref{tab:experiment_results} also empirically evaluates the performance of different design choices within \( d_{\text{Manifold}} \) as discussed in \Autoref{sec:graph_construction}. Results indicate that performance varies across datasets and embedding models.

\begin{figure*}
    \centering
    \includegraphics[width=\linewidth]{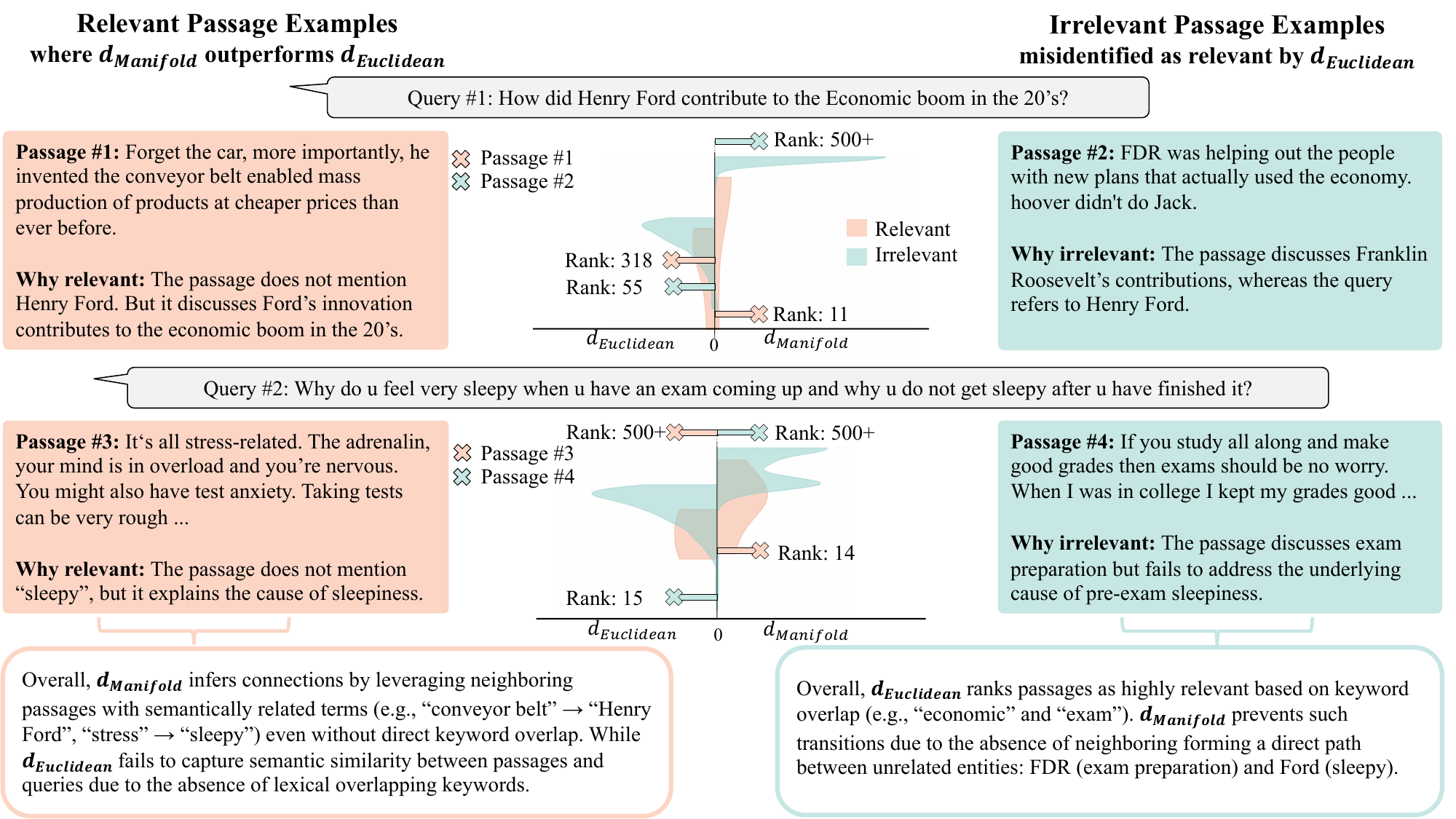}
\caption{Examples from ANTIQUE where MA-DPR outperforms DPR with \( d_{\text{Euclidean}} \) under \texttt{tas-b}. We present (i) relevant passages successfully retrieved by \( d_{\text{Manifold}} \) within the top 20 but missed by \( d_{\text{Euclidean}} \), and (ii) irrelevant passages that misidentified as relevant by \( d_{\text{Euclidean}} \). The Kernel Density illustrates the distribution of \( d_{\text{Euclidean}} \) and \( d_{\text{Manifold}} \) distances among the top 500 retrieved passages, categorized by ground truth relevance. \( d_{\text{Euclidean}} \) exhibits substantial overlap between relevant and irrelevant passages, failing to distinguish true relevance. In contrast, \( d_{\text{Manifold}} \) demonstrates clear separation.}
    \label{fig:RQ4}
\end{figure*}

For most OOD datasets, \(c^{\text{UC}}\) outperforms \(c^{\text{DC}}\).  We conjecture that \(c^{\text{UC}}\) prioritizes pure connectivity over raw distances in embedding space, which are often unreliable due to distortions in out-of-distribution settings. \(c^{\text{UC}}\) emphasizes the discrete transitions between neighboring passages rather than relying on possibly misleading distances. 
This is particularly beneficial when relevant passages are not directly similar to the query but lie along a chain of semantically related neighbors, which will be further discussed in RQ5 (cf. \Autoref{fig:RQ4}).
 
The performance between \( d^{\text{KNN}}_{\text{Euclidean}} \) and \( d^{\text{KNN}}_{\text{Spectral}} \) varies across embedding models. \( d^{\text{KNN}}_{\text{Euclidean}} \) performs better with \texttt{SciNCL} since \texttt{SciNCL} explicitly preserves local neighborhood structure through neighbor-aware contrastive learning \cite{ostendorff2022neighborhoodcontrastivelearningscientific} as its training objective, making \( d^{\text{KNN}}_{\text{Euclidean}} \) a strong fit for capturing local similarity.  In contrast, \texttt{tas-b} does not train with neighborhood-aware methods and hence \( d^{\text{KNN}}_{\text{Spectral}} \) methods recover neighborhood-aware embeddings that facilitate MA-DPR's manifold construction. 

These results suggest that dense embedding methods that are neighborhood-aware can use simple \( d^{\text{KNN}}_{\text{Euclidean}} \) distance for manifold construction, while \( d^{\text{KNN}}_{\text{Spectral}} \) is useful when the embeddings are not already optimized for neighborhood similarity.

\begin{figure}[h]
    \centering
    \includegraphics[width=\linewidth]{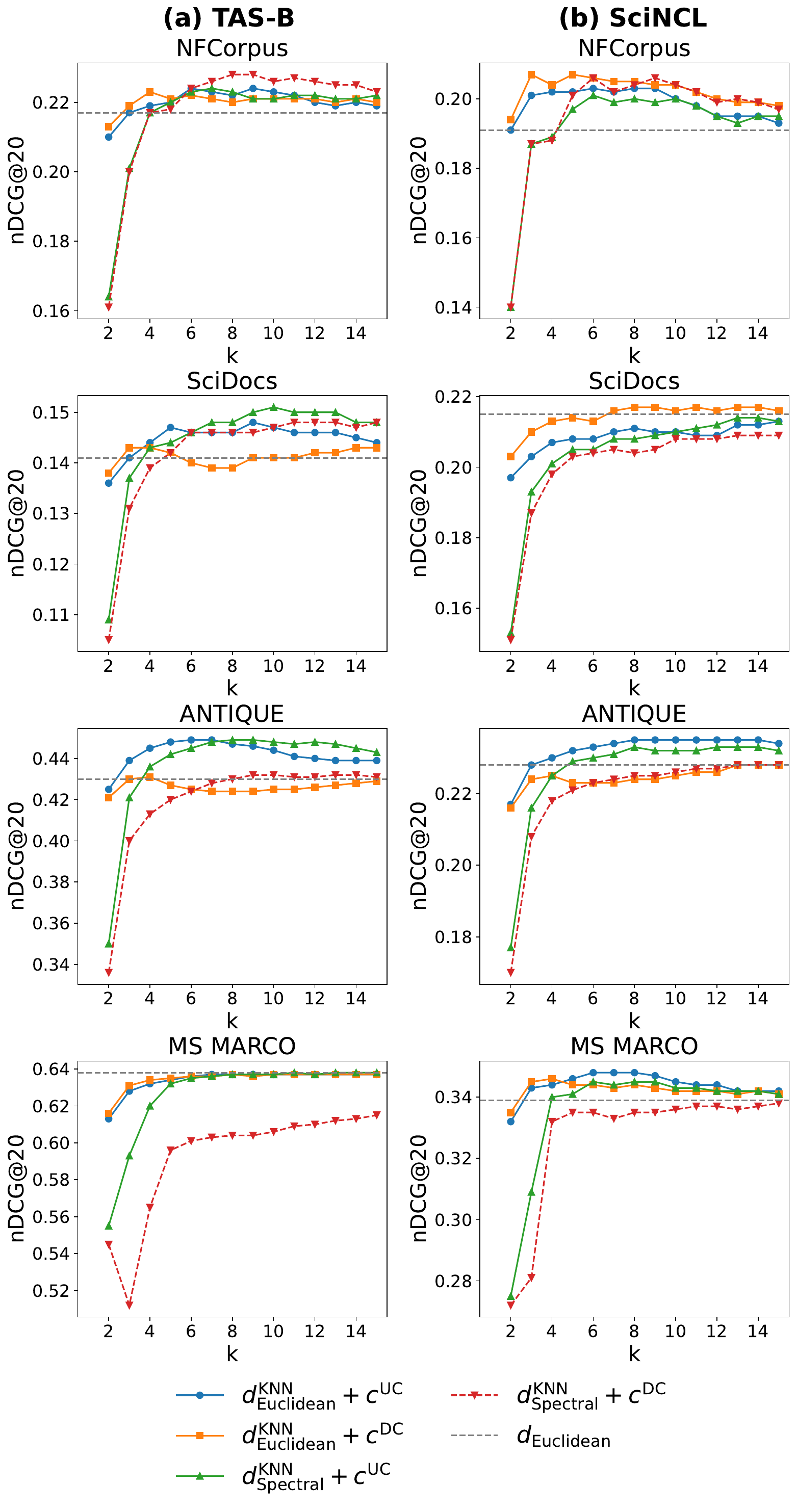}
\caption{Performance of MA-DPR in nDCG@20 across varying values of $K$ neighbors (x-axis).}
    \label{fig:ndcg20_vs_k_tasb_scincl_2x4}
\end{figure}

\paragraph{\textbf{RQ4} \textbf{Effect of K}: } 


\#$K$ neighbors controls the balance between sensitivity to noise and preservation of meaningful manifold structural relations, which can influence the effectiveness of MA-DPR and lead to performance variations. \Autoref{fig:ndcg20_vs_k_tasb_scincl_2x4} evaluates the performance of MA-DPR with $K \in  \{1,\ldots, 15\} $(in nDCG only, see \Autoref{sec:k} for full results). 

When $K$ is small (e.g., $K \leq 3$), performance remains substantially below other settings, suggesting that the graph becomes too sparse and retrieval is vulnerable to local noise. When $K$ is large (e.g., $K \geq 10$), performance converges toward that of DPR, indicating that the manifold approaches a highly smoothed structure and effectively reduces to a linear case that approximates \( d_{\text{Euclidean}} \), thereby diminishing the benefits of manifold modeling. Intermediate values of $K$ (around $K=6$--$8$) achieve the best trade-off, mitigating noise effects of individual documents while still preserving non-linear manifold structure, thus leading to consistently improved retrieval effectiveness across datasets.

\paragraph{\textbf{RQ5} \textbf{Reasons for Improvement}:} \Autoref{fig:RQ4} presents the distribution of \( d_{\text{Manifold}} \) and \( d_{\text{Euclidean}} \) across their respective top 500 retrieved passages for two example queries. In both cases, \( d_{\text{Euclidean}} \) fails to clearly distinguish relevant passages from irrelevant ones in the density plots (middle), as their distance distributions significantly overlap. In contrast, \( d_{\text{Manifold}} \) effectively separates relevant passages into a distinct range.

Further analysis of the context in these queries and passages (cf. text in \Autoref{fig:RQ4}) reveals cases where \( d_{\text{Euclidean}} \) has poor performance: (1) settings where relevant Passages 1 and 3 require reasoning and lack direct semantic overlap with the query, which \( d_{\text{Euclidean}} \) fails to identify as relevant; and (2) settings such as irrelevant Passage 4 that contains partially overlapping keywords but different contextual meanings, or Passage 2 where two historical figures have a strong economic association.  Such misleading similarities cause \( d_{\text{Euclidean}} \) to erroneously rank them as relevant.

In contrast, \( d_{\text{Manifold}} \) remains effective in all cases of \Autoref{fig:RQ4}: (1)~For relevant Passages 1 and 3, \( d_{\text{Manifold}} \) leverages neighboring passages to provide crucial missing context that bridges the query and passage in the absence of direct similarity.  (2) For irrelevant Passages 2 and 4 with misleading lexical or semantic overlap, the lack of semantically similar neighboring passages precludes \( d_{\text{Manifold}} \) from small distances, and hence relevance to the query. 

We believe these anecdotal examples show how MA-DPR can leverage information in related neighboring passages, unlike DPR that does not consider neighborhood structure during retrieval.
\section{Conclusion}
With the aim to better capturing relevance in DPR via distance on the subdimensional non-linear manifold of query and passage embeddings, we introduced novel distance metrics for DPR to leverage the underlying manifold structure of embeddings using a graph-based representation. 

With a one-time computational cost for graph construction and comparable online query inference cost to standard DPR, our proposed MA-DPR is able to exploit the manifold structure of embedding space and achieves up to a 26\% improvement over DPR using traditional Euclidean and cosine distances, particularly on OOD datasets.

By leveraging the context from neighboring passages, manifold-aware DPR demonstrates effectiveness for queries and passages lacking overlapping keywords.
These findings suggest that manifold-aware distance can significantly enhance DPR performance and that 
search over the implicit manifold of data can help overcome deficiencies in embedding training for OOD settings.
\section{Limitations}
While the proposed MA-DPR effectively models non-linear structures in embedding space, several limitations remain. First, the current approach relies on a complete KNN graph constructed offline, which may pose scalability challenges as the number of passages grows. 
However, (approximate) nearest neighbor methods such as  SCANN~\citep{scann}, SOAR~\citep{soar}, or FAISS~\citep{FAISS}, combined with incremental graph construction techniques can build the graph on-demand and limit the space and search complexity of the graph to the subset of top-ranked documents explored during the manifold search.
Second, the graph edge weights are derived from unsupervised distance metrics, which may not always align with relevance signals in retrieval tasks. Incorporating supervised signals to learn edge weights can further refine the manifold representation and improve retrieval effectiveness.

\bibliography{main}

\clearpage
\newpage
\appendix
\section{Failure Analysis}

While MA-DPR demonstrates strong performance overall, we identify scenarios in which the manifold-aware distance may lead to degraded retrieval effectiveness. A key limitation arises when relevant passages contain highly specific or technical terminology that is not well connected to other passages in the constructed manifold graph. Since MA-DPR propagates relevance through graph connectivity, such isolated passages may be overlooked and ranked much lower than in direct distance-based retrieval.

\paragraph{Example.} 
Consider the query:
\begin{quote}
    \textit{``Stochastic Variational Deep Kernel Learning''}
\end{quote}
with two relevant passages:
\begin{itemize}
    \item Passage \#1: \textit{``In this paper, we introduce deep Gaussian process (GP) models. Deep GPs are \dots''}
    \item Passage \#2: \textit{``We develop a scalable deep non-parametric generative model by augmenting deep Gaussian processes with a recognition model. Inference is performed in \dots''}
\end{itemize}

Under Euclidean distance retrieval, these passages are ranked at positions 12 and 37. By contrast, MA-DPR ranks the same passages much lower at positions 2296 and 2586. The substantial drop in rank highlights a structural challenge: these passages are weakly connected within the manifold graph, as reflected by their relatively low number of edges compared to the average node degree. As a result, the manifold-based distance disperses their relevance across unrelated neighbors, diminishing their retrieval scores. This may also help explain MA-DPR's reduced effectiveness on scientifically dense datasets such as SCIDOCS, where many relevant passages are highly technical and sparsely connected.

\section{Baseline Methods and Settings}
\label{sec:baseline}

\subsection{Euclidean and cosine Distance Baselines}

Let \( \mathbf{e}_q \in \mathbb{R}^D \) denote the query embedding and \( \mathbf{e}_p^{(i)} \in \mathbb{R}^D \) denote the embedding of passage \( p^{(i)} \). Distance-based retrieval uses similarity metrics directly in the original embedding space without additional transformation.

\paragraph{Euclidean Distance}
\begin{equation}
    d_{\text{Euclidean}}(\mathbf{e}_q, \mathbf{e}_p^{(i)}) 
    = \|\mathbf{e}_q - \mathbf{e}_p^{(i)}\|_2
\end{equation}

\paragraph{Cosine Distance}
\begin{equation}
    d_{\text{Cosine}}(\mathbf{e}_q, \mathbf{e}_p^{(i)}) 
    = 1 - \frac{\mathbf{e}_q \cdot \mathbf{e}_p^{(i)}}{\|\mathbf{e}_q\|_2 \|\mathbf{e}_p^{(i)}\|_2}
\end{equation}

These two metrics serve as standard baselines for dense passage retrieval (DPR).

\subsection{Unsupervised Projection Baselines}

We further consider mapping-based baselines, where embeddings are first mapped into a lower-dimensional space before retrieval. We first examine unsupervised projection methods, which rely solely on the structure of the embedding distribution without relevance labels.

\paragraph{Principal Component Analysis (PCA)} 
PCA learns a linear projection \( W \in \mathbb{R}^{D \times d} \) such that
\begin{equation}
    \mathbf{z}_p^{(i)} = W^\top \mathbf{e}_p^{(i)}, \quad \mathbf{z}_q = W^\top \mathbf{e}_q,
\end{equation}
where \( d < D \) is the target dimension. The projection matrix \( W \) is chosen to maximize the variance of the projected embeddings, equivalently solving
\begin{equation}
    W = \arg\max_{W^\top W = I_d} \; \mathrm{Tr}\left(W^\top \Sigma W\right),
\end{equation}
where \( \Sigma \) denotes the empirical covariance matrix of the embeddings.

\paragraph{Kernel Principal Component Analysis (KernelPCA)} 
KernelPCA extends PCA by mapping embeddings into a reproducing kernel Hilbert space (RKHS) via a non-linear feature map \( \phi(\cdot) \), and then performing PCA in that space. The kernel function is defined as
\begin{equation}
    K(x, x') = \langle \phi(x), \phi(x') \rangle.
\end{equation}
The optimization objective is identical in form to PCA, namely to maximize variance in the projected space:
\begin{equation}
    U = \arg\max_{U^\top U = I_d} \; \mathrm{Tr}\left(U^\top K U\right),
\end{equation}
where \( K \) is the centered kernel matrix and \( U \) contains the top \( d \) eigenvectors. 

We consider the following kernels:
\begin{itemize}
    \item \textbf{Linear}: \( K(x, x') = x^\top x' \)
    \item \textbf{Polynomial}: \( K(x, x') = (\gamma x^\top x' + c_0)^{d_k} \)
    \item \textbf{RBF}: \( K(x, x') = \exp(-\gamma \|x - x'\|^2) \)
\end{itemize}

For these unsupervised projections, we vary the target dimension \( M \in \{100, 500\} \).

We use the \texttt{sklearn.decomposition.KernelPCA} implementation from scikit-learn. Kernel hyperparameters (\( \gamma, c_0, d_k \)) are left at their default values unless otherwise specified. Full results are reported in \Autoref{tab:dr_models_results_combined}.

\subsection{Supervised Metric learning Baselines}

In addition to unsupervised projections, we evaluate supervised projection methods that leverage query–document relevance labels to learn task-specific transformations of the embedding space.

\paragraph{Neighborhood Components Analysis (NCA)} 
NCA learns a linear projection \( W \in \mathbb{R}^{D \times d} \) that minimizes the expected classification error under a stochastic nearest-neighbor classifier. The probability of assigning query \( q \) to passage \( p^{(i)} \) is defined as
\begin{equation}
    P(i \mid q) = 
    \frac{\exp\!\left(-\|W^\top \mathbf{e}_q - W^\top \mathbf{e}_p^{(i)}\|_2^2\right)}
         {\sum_j \exp\!\left(-\|W^\top \mathbf{e}_q - W^\top \mathbf{e}_p^{(j)}\|_2^2\right)}.
\end{equation}

Let \( \mathcal{R}(q) \) denote the set of relevant passages for query \( q \). The learning objective of NCA is to maximize the probability that each query is correctly assigned to one of its relevant passages:
\begin{equation}
    \max_W \; \sum_{q} \sum_{i \in \mathcal{R}(q)} P(i \mid q).
\end{equation}

Similar to the unsupervised setting, we vary the projection dimension \( M \in \{100, 500\} \). 

We use the \texttt{sklearn.\allowbreak neighbors.\allowbreak NeighborhoodComponentsAnalysis}
 implementation from scikit-learn’s metric learning module. For MS MARCO, NFCorpus, and ANTIQUE, which provide both training and test sets, we train the mapping on the training set and evaluate performance on the test set. We do not report results for SciDocs, as it does not provide a train/test split. During training, we run the algorithm for 100 iterations (using the default L-BFGS solver). Full results are reported in \Autoref{tab:dr_models_results_combined}.

\begin{table*}
\centering
\caption{Retrieval performance with mapping-based baselines across four datasets for embedding dimensions $M \in \{100,500\}$. Panel A uses \texttt{tas-b} embeddings; Panel B uses \texttt{SciNCL}. A slash (/) indicates results not reported.}
\resizebox{\textwidth}{!}{
\setlength{\tabcolsep}{3pt}
\begin{threeparttable}
\begin{tabular}{lccc|ccc|ccc|ccc}
\toprule
 & \multicolumn{3}{c}{\textbf{NFCorpus}} 
 & \multicolumn{3}{c}{\textbf{SciDocs}}
 & \multicolumn{3}{c}{\textbf{ANTIQUE}}
 & \multicolumn{3}{c}{\textbf{MS MARCO}} \\
\cmidrule(lr){2-4} \cmidrule(lr){5-7} \cmidrule(lr){8-10} \cmidrule(lr){11-13}
 & R@20 & mAP@20 & nDCG@20 & R@20 & mAP@20 & nDCG@20 & R@20 & mAP@20 & nDCG@20 & R@20 & mAP@20 & nDCG@20 \\
\midrule
\multicolumn{13}{c}{\textbf{Panel A: \texttt{tas-b}}} \\
\midrule
\multicolumn{13}{c}{\centering \textbf{Linear PCA}} \\
$M=100$ & 0.086 & 0.076 & 0.147 & 0.131 & 0.053 & 0.105 & 0.174 & 0.101 & 0.184 & 0.944 & 0.532 & 0.634 \\\addlinespace[2pt]
$M=500$ & 0.123 & 0.078 & 0.201 & 0.175 & 0.076 & 0.146 & 0.236 & 0.145 & 0.246 & 0.951 & 0.550 & 0.650 \\
\midrule
\multicolumn{13}{c}{\centering \textbf{Kernel PCA (Quadratic)}} \\
$M=100$ & 0.087 & 0.076 & 0.147 & 0.130 & 0.053 & 0.105 & 0.163 & 0.092 & 0.171 & 0.944 & 0.532 & 0.634 \\\addlinespace[2pt]
$M=500$ & 0.121 & 0.076 & 0.200 & 0.175 & 0.076 & 0.146 & 0.233 & 0.143 & 0.243 & 0.951 & 0.551 & 0.651 \\
\midrule
\multicolumn{13}{c}{\centering \textbf{Kernel PCA (RBF)}} \\
$M=100$ & 0.088 & 0.078 & 0.149 & 0.131 & 0.053 & 0.106 & 0.176 & 0.103 & 0.186 & 0.944 & 0.531 & 0.633 \\\addlinespace[2pt]
$M=500$ & 0.123 & 0.078 & 0.201 & 0.176 & 0.077 & 0.146 & 0.240 & 0.147 & 0.248 & 0.951 & 0.552 & 0.651 \\
\midrule
\multicolumn{13}{c}{\centering \textbf{Metric Learning (NCA)}} \\
$M=100$ & 0.079 & 0.071 & 0.133 & / & / & / & 0.250 & 0.165 & 0.276 & 0.951 & 0.535 & 0.639 \\\addlinespace[2pt]
$M=500$ & 0.138 & 0.081 & 0.220 & / & / & / & 0.263 & 0.171 & 0.282 & 0.951 & 0.535 & 0.639 \\
\midrule
\multicolumn{13}{c}{\textbf{Panel B: \texttt{SciNCL}}} \\
\midrule
\multicolumn{13}{c}{\centering \textbf{Linear PCA}} \\
$M=100$ & 0.112 & 0.069 & 0.178 & 0.250 & 0.108 & 0.199 & 0.105 & 0.054 & 0.108 & 0.625 & 0.248 & 0.338 \\\addlinespace[2pt]
$M=500$ & 0.115 & 0.070 & 0.183 & 0.260 & 0.112 & 0.206 & 0.111 & 0.060 & 0.116 & 0.637 & 0.253 & 0.346 \\
\midrule
\multicolumn{13}{c}{\centering \textbf{Kernel PCA (Quadratic)}} \\
$M=100$ & 0.110 & 0.068 & 0.177 & 0.251 & 0.108 & 0.200 & 0.095 & 0.049 & 0.098 & 0.623 & 0.250 & 0.341 \\\addlinespace[2pt]
$M=500$ & 0.115 & 0.069 & 0.183 & 0.260 & 0.112 & 0.206 & 0.104 & 0.055 & 0.108 & 0.630 & 0.252 & 0.344 \\
\midrule
\multicolumn{13}{c}{\centering \textbf{Kernel PCA (RBF)}} \\
$M=100$ & 0.110 & 0.068 & 0.176 & 0.251 & 0.107 & 0.199 & 0.104 & 0.053 & 0.104 & 0.628 & 0.251 & 0.342 \\\addlinespace[2pt]
$M=500$ & 0.115 & 0.070 & 0.182 & 0.259 & 0.111 & 0.205 & 0.109 & 0.059 & 0.115 & 0.639 & 0.254 & 0.346 \\
\midrule
\multicolumn{13}{c}{\centering \textbf{Metric Learning (NCA)}} \\
$M=100$ & 0.067 & 0.066 & 0.119 & / & / & / & 0.131 & 0.074 & 0.141 & 0.626 & 0.250 & 0.341 \\\addlinespace[2pt]
$M=500$ & 0.100 & 0.078 & 0.157 & / & / & / & 0.114 & 0.065 & 0.124 & 0.605 & 0.228 & 0.314 \\
\bottomrule
\end{tabular}
\end{threeparttable}}
\label{tab:dr_models_results_combined}
\end{table*}

\section{Impact of Normalization}
\label{sec:norm}
In this section, we examine the effect of embedding normalization on the performance of our proposed MA-DPR.

As shown in \Autoref{tab:experiment_results_unnorm}, we observe that removing normalization leads to a performance drop for both the baseline \( d_{\text{Euclidean}} \) and \( d_{\text{Manifold}} \), with the degradation occurring at a similar level. Importantly, the performance gap between \( d_{\text{Manifold}} \) and the baseline remains consistent, suggesting that MA-DPR is robust to whether embeddings are normalized or not.

\begin{table*}[htbp]
\centering
\caption{Performance of \( d_{\text{Manifold}} \) across design choices on four datasets. 
We use original embeddings \textbf{without normalization}.}
\scalebox{0.73}{
\setlength{\tabcolsep}{3pt} 
\begin{tabular}{lccc|ccc|ccc|ccc}
\toprule
 & \multicolumn{3}{c}{\textbf{NFCorpus}} 
 & \multicolumn{3}{c}{\textbf{SciDocs}}
 & \multicolumn{3}{c}{\textbf{ANTIQUE}} 
 & \multicolumn{3}{c}{\textbf{MS MARCO}} \\
\cmidrule(lr){2-4} \cmidrule(lr){5-7} \cmidrule(lr){8-10} \cmidrule(lr){11-13}
 & R@20 & mAP@20 & nDCG@20 & R@20 & mAP@20 & nDCG@20 & R@20 & mAP@20 & nDCG@20 & R@20 & mAP@20 & nDCG@20 \\
\midrule
\multicolumn{13}{c}{\centering msmarco-distilbert-base-tas-b} \\
\midrule
$d_{\text{Euclidean}}$ & 0.110 & 0.065 & 0.185 & 0.155 & 0.065 & 0.127 & 0.409 & 0.283 & 0.412 & \textbf{0.945} & \textbf{0.526} & \textbf{0.630} \\\addlinespace[5pt]
$d^{\text{KNN}}_{\text{Euclidean}}$+ $c^{\text{UC}}$ & 0.123 & 0.070 & 0.196 & 0.166 & 0.068 & 0.131 & \textbf{0.451} & 0.298 & 0.431 & 0.944 & \textbf{0.526} & \textbf{0.630} \\\addlinespace[2pt]
$d^{\text{KNN}}_{\text{Euclidean}}$+$c^{\text{DC}}$ & 0.111 & 0.069 & 0.193 & 0.148 & 0.065 & 0.124 & 0.389 & 0.282 & 0.402 & 0.940 & 0.525 & 0.629 \\\addlinespace[2pt]
$d^{\text{KNN}}_{\text{Spectral}}$+$c^{\text{UC}}$ & \textbf{0.129} & 0.072 & \textbf{0.199} & \textbf{0.175} & \textbf{0.070} & \textbf{0.136} & 0.448 & \textbf{0.303} & \textbf{0.434} & 0.944 & \textbf{0.526} & 0.629 \\\addlinespace[2pt] 
$d^{\text{KNN}}_{\text{Spectral}}$+$c^{\text{DC}}$ & 0.126 & \textbf{0.074} & 0.197 & 0.159 & 0.068 & 0.126 & 0.385 & 0.279 & 0.387 & 0.928 & 0.483 & 0.601 \\
\midrule
\multicolumn{13}{c}{\centering SciNCL}  \\ 
\midrule
$d_{\text{Euclidean}}$ & 0.120 & 0.073 & 0.190 & 0.279 & 0.117 & 0.217 & 0.238 & 0.138 & 0.227 & 0.616 & 0.250 & 0.338 \\\addlinespace[5pt]
$d^{\text{KNN}}_{\text{Euclidean}}$+$c^{\text{UC}}$ & \textbf{0.131} & \textbf{0.080} & 0.202 & 0.269 & 0.117 & 0.212 & \textbf{0.249} & \textbf{0.144} & \textbf{0.233} & \textbf{0.652} & \textbf{0.254} & \textbf{0.348} \\\addlinespace[2pt]
$d^{\text{KNN}}_{\text{Euclidean}}$+$c^{\text{DC}}$ & 0.130 & \textbf{0.080} & \textbf{0.204} & \textbf{0.281} & \textbf{0.119} & \textbf{0.218} & 0.224 & 0.138 & 0.221 & 0.630 & 0.252 & 0.342 \\\addlinespace[2pt]
$d^{\text{KNN}}_{\text{Spectral}}$+$c^{\text{UC}}$ & \textbf{0.131} & 0.079 & 0.201 & 0.260 & 0.116 & 0.209 & 0.243 & 0.143 & 0.231 & 0.636 & 0.253 & 0.344 \\\addlinespace[2pt]
$d^{\text{KNN}}_{\text{Spectral}}$+$c^{\text{DC}}$ & 0.117 & \textbf{0.080} & 0.196 & 0.226 & 0.099 & 0.180 & 0.203 & 0.126 & 0.200 & 0.577 & 0.229 & 0.311 \\
\bottomrule
\end{tabular}
\label{tab:experiment_results_unnorm}
}
\end{table*}

\section{Additional Embedding}

We additionally report results using \texttt{msmarco-distilbert-dot-v5}~\citep{reimers-2019-sentence-bert} to provide further empirical support for our method, trained on MS MARCO. The results are consistent with our main findings and align with the discussions presented in the paper (cf. \Autoref{tab:experiment_resultsz_v5}).

\begin{table*}
\centering
\caption{Performance of \( d_{\text{Manifold}} \) across design choices on four datasets with \textbf{\texttt{dot-v5}} embeddings. 
The best results are highlighted in \textbf{bold}. An asterisk (*) denotes a statistically significant improvement of MA-DPR over DPR baselines (paired \( t \)-test, \( p < 0.05 \)). A slash (/) indicates results not reported. We normalize \texttt{tas-b} and \texttt{SciNCL} embeddings so that \( d_{\text{Euclidean}} \) and \( d_{\text{Cosine}} \) produce identical rankings.}
\scalebox{0.73}{
\setlength{\tabcolsep}{3pt} 
\begin{tabular}{lccc|ccc|ccc|ccc}
\toprule
 & \multicolumn{3}{c}{\textbf{NFCorpus}} 
 & \multicolumn{3}{c}{\textbf{SciDocs}}
 & \multicolumn{3}{c}{\textbf{ANTIQUE}} 
 & \multicolumn{3}{c}{\textbf{MS MARCO}} \\
\cmidrule(lr){2-4} \cmidrule(lr){5-7} \cmidrule(lr){8-10} \cmidrule(lr){11-13}
 & R@20 & mAP@20 & nDCG@20 & R@20 & mAP@20 & nDCG@20 & R@20 & mAP@20 & nDCG@20 & R@20 & mAP@20 & nDCG@20 \\
\midrule
    \multicolumn{13}{c}{\centering msmarco-bert-base-dot-v5} \\
    \midrule
    $d_{\text{Euclidean}}$ ($d_{\text{Cosine}}$) & 0.132 & 0.071 & 0.192 & 0.168 & 0.070 & 0.138 & 0.382 & 0.251 & 0.377 & 0.937 & \textbf{0.529} & 0.631 \\\addlinespace[5pt]
   $d^{\text{KNN}}_{\text{Euclidean}}$+ $c^{\text{UC}}$ & 0.133 & 0.077 & 0.202 & 0.181 & 0.079 & 0.147 & 0.423 & 0.268 & 0.398 & 0.942 & \textbf{0.529} & 0.632 \\\addlinespace[2pt]
    $d^{\text{KNN}}_{\text{Euclidean}}$+$c^{\text{DC}}$ & 0.133 & 0.078 & 0.204 & 0.181 & 0.079 & 0.147 & 0.365 & 0.253 & 0.372 & 0.936 & \textbf{0.529} & 0.631 \\\addlinespace[2pt]
    $d^{\text{KNN}}_{\text{Spectral}}$+$c^{\text{UC}}$ & 0.136 & 0.077 & 0.203 & \textbf{0.183}* & \textbf{0.080} & \textbf{0.148} & \textbf{0.427}* & 0.275 & \textbf{0.404}* & \textbf{0.943} & \textbf{0.529} & \textbf{0.633} \\\addlinespace[2pt] 
  $d^{\text{KNN}}_{\text{Spectral}}$+$c^{\text{DC}}$ & \textbf{0.139} & \textbf{0.078} & \textbf{0.207}* & 0.182 & \textbf{0.080} & \textbf{0.148} & 0.411 & \textbf{0.277}* & 0.399 & 0.929 & 0.499 & 0.604 \\
\bottomrule
\end{tabular}\label{tab:experiment_resultsz_v5}
}
\end{table*}

\section{Runtime Simulation}
\label{sec:runtime}
In the runtime simulation, we evaluate the per-query latency of DPR and MA-DPR across a grid of embedding dimensions \(D\) and graph neighborhood sizes \(K\). \Autoref{tab:runtime_madpr_by_k_with_dpr} shows that the runtime of both methods scales approximately linearly with \(D\), which is consistent with their theoretical complexities. In real-world retrieval scenarios, \(K\) is typically small (e.g., fewer than 10), so the overall runtime is dominated by \(D\) and the choice of \(K\) has little influence. Importantly, the runtime overhead between MA-DPR and DPR remains consistently low, typically 2-4 milliseconds per query. This indicates that the additional manifold-aware shortest-path computation introduces negligible latency, while preserving the efficiency of standard dense retrieval.  

We also report graph construction times in \Autoref{tab:graph_construction_results}. Unlike query latency, graph construction exhibits substantial cost, increasing with the embedding dimension from tens of seconds at \(D=32\) to several minutes at \(D=1024\). However, this procedure is performed only once as a preprocessing step, and the resulting graph can then be reused for all subsequent queries. As such, although graph construction dominates the total wall-clock time during setup, it does not affect inference-time efficiency and can be amortized across large-scale retrieval workloads.

\begin{table*}[htbp]
\centering
\caption{Per-query inference runtimes over 100$k$ passages. Columns list MA-DPR runtimes (ms) for each \(K\); the final column reports the DPR mean (ms) across \(K\) at each dimension \(D\).}
\resizebox{\textwidth}{!}{%
\begin{tabular}{l c c c c c c c c c c c c c c c}
\toprule
\multirow{2}{*}{\textbf{Dimension (D)}} & \multicolumn{14}{c}{\textbf{MA-DPR Runtime (ms) by \textit{K}}} & \multirow{2}{*}{\textbf{DPR Mean (ms)}} \\
\cmidrule(lr){2-15}
 & \textbf{2} & \textbf{3} & \textbf{4} & \textbf{5} & \textbf{6} & \textbf{7} & \textbf{8} & \textbf{9} & \textbf{10} & \textbf{11} & \textbf{12} & \textbf{13} & \textbf{14} & \textbf{15} & \\
\midrule
32 & 7.48 & 7.71 & 8.01 & 7.34 & 8.02 & 8.34 & 8.33 & 7.84 & 8.30 & 7.40 & 8.04 & 8.26 & 7.63 & 8.14 & 5.05 \\
64 & 12.03 & 11.60 & 11.61 & 11.28 & 11.72 & 12.19 & 11.76 & 11.84 & 12.01 & 11.97 & 12.08 & 12.15 & 12.22 & 11.75 & 9.16 \\
128 & 27.20 & 27.13 & 27.60 & 27.80 & 27.28 & 27.65 & 28.32 & 27.66 & 27.49 & 27.98 & 27.82 & 27.65 & 27.33 & 28.27 & 25.00 \\
256 & 58.36 & 58.90 & 58.69 & 58.12 & 58.84 & 59.34 & 58.85 & 58.30 & 59.00 & 59.38 & 59.19 & 58.98 & 59.87 & 59.92 & 54.55 \\
512 & 110.24 & 110.52 & 109.87 & 109.57 & 110.26 & 109.98 & 111.11 & 110.41 & 110.40 & 109.94 & 110.76 & 110.40 & 111.13 & 111.06 & 107.62 \\
1024 & 205.87 & 206.13 & 205.19 & 205.48 & 205.87 & 205.40 & 205.32 & 205.50 & 205.84 & 206.08 & 206.01 & 206.17 & 205.67 & 205.83 & 202.35 \\
\bottomrule
\end{tabular}
}
\label{tab:runtime_madpr_by_k_with_dpr}
\end{table*}

\begin{table*}[htbp]
\centering
\caption{Graph construction time (s) over 100$k$ passages for each $(D,K)$ configuration.}
\resizebox{\textwidth}{!}{%
\begin{tabular}{lcccccccccccccc}
\toprule
\textbf{Dimension (D)} & \textbf{K=2} & \textbf{K=3} & \textbf{K=4} & \textbf{K=5} & \textbf{K=6} & \textbf{K=7} & \textbf{K=8} & \textbf{K=9} & \textbf{K=10} & \textbf{K=11} & \textbf{K=12} & \textbf{K=13} & \textbf{K=14} & \textbf{K=15} \\
\midrule
32 & 19.67 & 19.35 & 19.99 & 20.02 & 19.86 & 20.57 & 20.20 & 20.69 & 20.94 & 21.03 & 21.15 & 21.37 & 21.94 & 22.32 \\
64 & 23.91 & 24.05 & 24.24 & 24.76 & 24.89 & 24.70 & 25.08 & 25.08 & 25.75 & 25.63 & 26.24 & 26.62 & 26.40 & 26.51 \\
128 & 33.28 & 34.10 & 34.17 & 34.02 & 34.39 & 34.61 & 34.76 & 35.10 & 35.24 & 35.24 & 35.87 & 35.79 & 35.99 & 36.45 \\
256 & 53.14 & 53.86 & 53.38 & 54.38 & 54.41 & 54.76 & 54.51 & 54.47 & 55.08 & 55.45 & 55.38 & 55.93 & 56.00 & 55.92 \\
512 & 96.53 & 96.35 & 95.94 & 96.31 & 96.02 & 97.45 & 97.60 & 96.83 & 96.91 & 97.66 & 97.60 & 97.60 & 98.62 & 98.35 \\
1024 & 177.45 & 178.60 & 178.38 & 177.29 & 178.91 & 178.80 & 179.67 & 181.08 & 179.31 & 178.87 & 180.41 & 180.08 & 181.65 & 181.96 \\
\bottomrule
\end{tabular}
}
\label{tab:graph_construction_results}
\end{table*}

\section{Impact of Hyperparameters}
\label{sec:k}
\begin{figure*}[htbp]
    \centering
    \includegraphics[width=\textwidth]{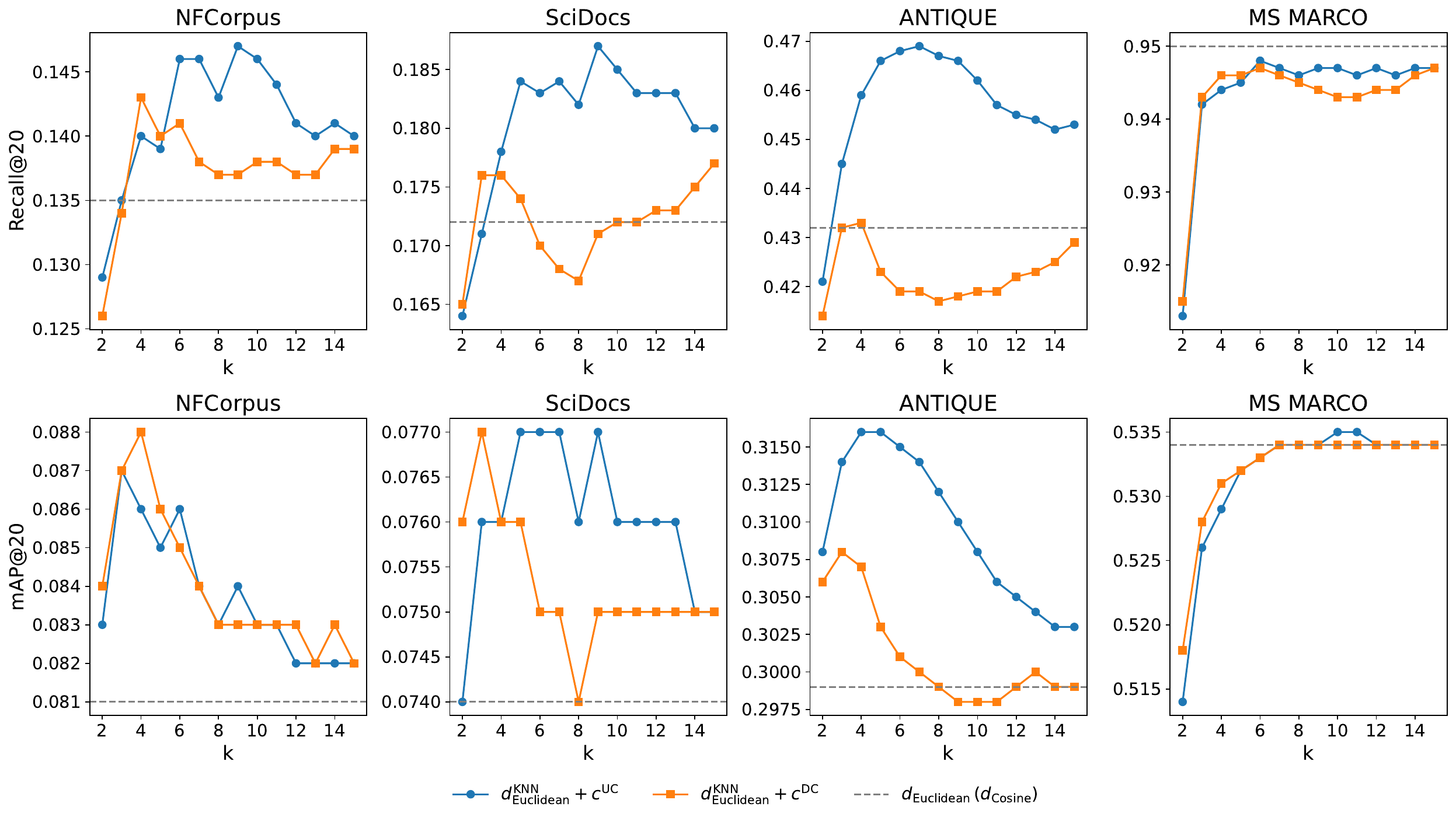}
    \caption{Recall@20 and mAP@20 comparison of $d^{\text{KNN}}_{\text{Euclidean}}$ with \texttt{tas-b} across varying $K$.}
    \label{fig:recall_and_map_2x4_combined_tasb}
\end{figure*}

\begin{figure*}[htbp]
    \centering
    \includegraphics[width=\textwidth]{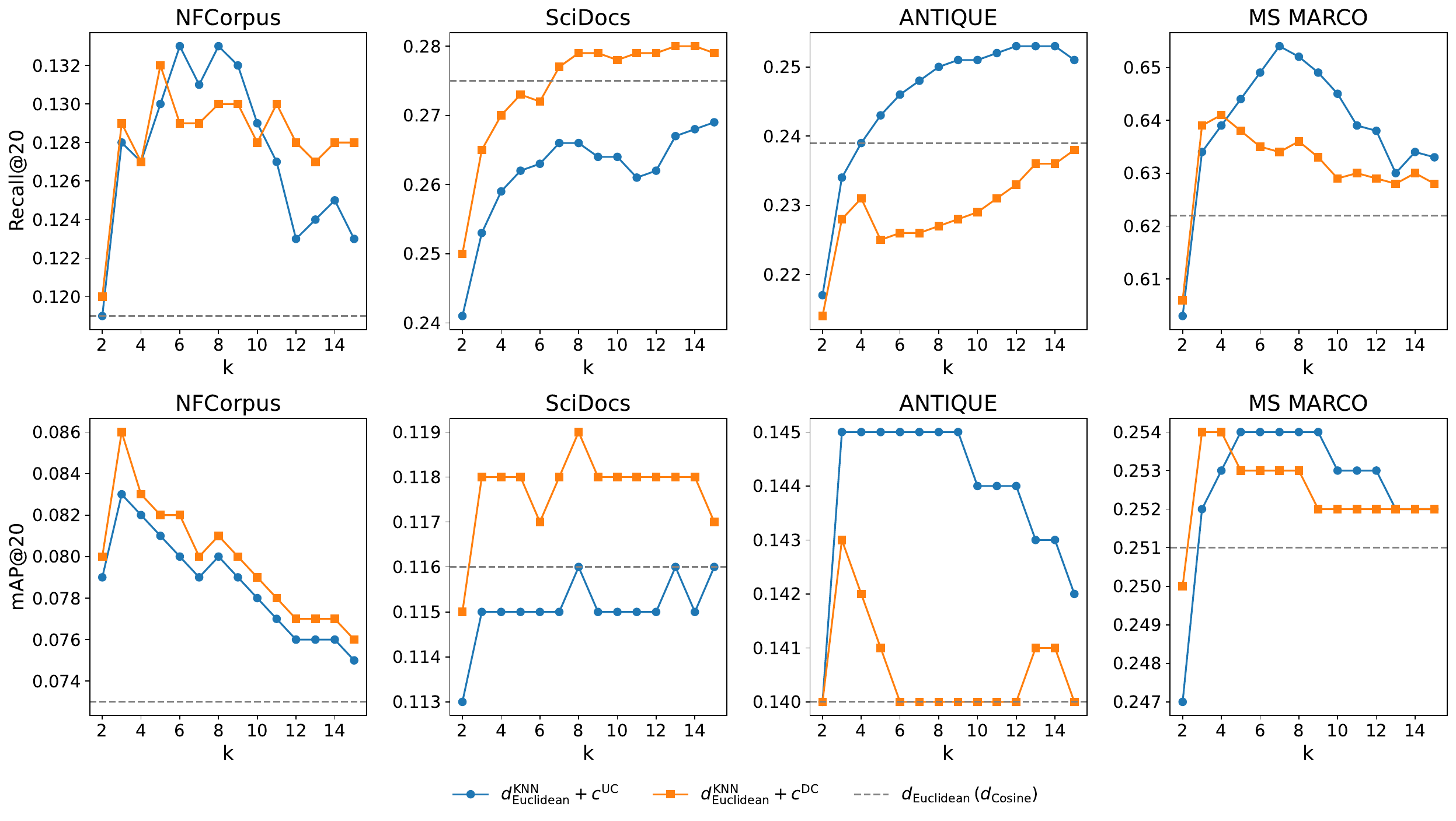}
    \caption{Recall@20 and mAP@20 comparison of $d^{\text{KNN}}_{\text{Euclidean}}$ with SciNCL across varying $K$.}
    \label{fig:recall_and_map_2x4_combined_sincil}
\end{figure*}

\begin{figure*}[htbp]
    \centering
    \includegraphics[width=\textwidth]{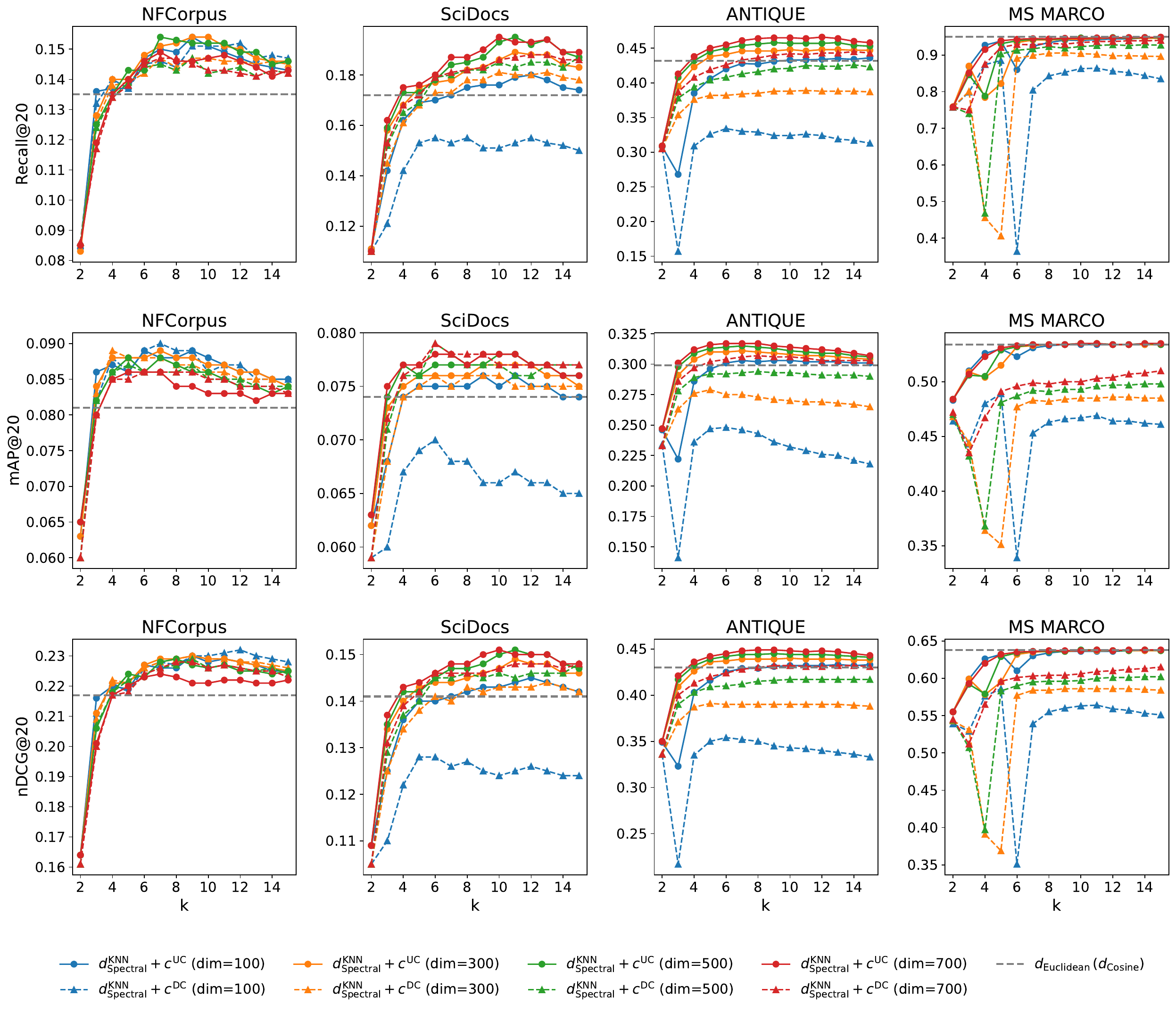}
    \caption{Recall@20, mAP@20 and nDCG@20 comparison of $d^{\text{KNN}}_{\text{Spectral}}$ with \texttt{tas-b} across varying $K$ and $M$.}
    \label{fig:spectral_Recall_map_nDCG_3x4_tasb}
\end{figure*}

\begin{figure*}[htbp]
    \centering
    \includegraphics[width=\textwidth]{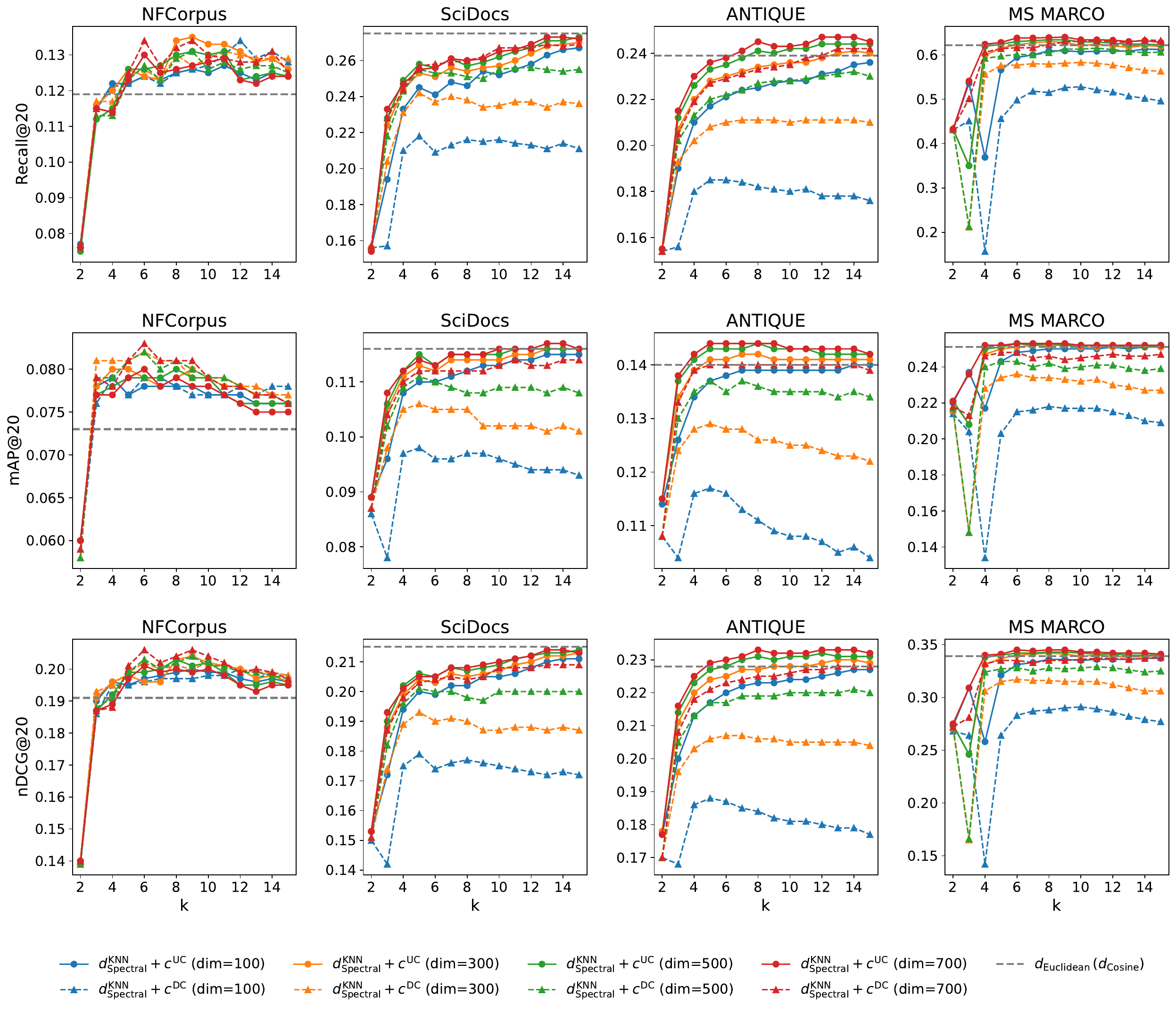}
    \caption{Recall@20, mAP@20 and nDCG@20 comparison of $d^{\text{KNN}}_{\text{Spectral}}$ with SciNCL across varying $K$ and $M$.}
    \label{fig:spectral_Recall_map_nDCG_3x4_scincl}
\end{figure*}

We present the full results on the effect of \(K\) for \( d^{\text{KNN}}_{\text{Euclidean}} \) using mAP and Recall in \Autoref{fig:recall_and_map_2x4_combined_tasb}, \Autoref{fig:recall_and_map_2x4_combined_sincil}, \Autoref{fig:spectral_Recall_map_nDCG_3x4_tasb}, and \Autoref{fig:spectral_Recall_map_nDCG_3x4_scincl}. The observed trend is consistent with the nDCG results discussed in RQ4 of \Autoref{sec:experiments}: performance is poor when \(K\) is small, improves at intermediate values of \(K\), and eventually collapses to the linear case where \( d_{\text{Manifold}} \approx d_{\text{Euclidean}} \).  

We also report the effect of the spectral dimension \(M\) on \( d^{\text{KNN}}_{\text{Spectral}} \) in \Autoref{fig:spectral_Recall_map_nDCG_3x4_tasb} and \Autoref{fig:spectral_Recall_map_nDCG_3x4_scincl}. The results show that performance is low at small values of \(M\) (e.g., 100 and 300) but converges at higher values (e.g., 500 and 700), indicating that sufficient spectral dimensionality is necessary to capture the underlying manifold structure.

\end{document}